\documentclass{vgtc}                          

\usepackage{mathptmx}
\usepackage{graphicx}
\usepackage{times,subfig}
\usepackage{balance}

\onlineid{0}

\vgtccategory{Research}

\vgtcinsertpkg

\title{Scalable Rendering of variable Density Point Cloud Data}

\author{Priyadarshini Kumari\thanks{priyadarshini.k@iitb.ac.in}\\ %
        \scriptsize Department of Electrical Engineering,\\ IIT Bombay %
\and Sreeni K.G\thanks{sreenikg79@gmail.com}\\ %
     \scriptsize College of Engineering,\\ Thiruvananathapuram %
\and Subhasis Chaudhuri\thanks{sc@ee.iitb.ac.in}\\ %
     \scriptsize Department of Electrical Engineering,\\ IIT Bombay 
   }

\abstract{In this paper, we present a novel proxy based method of adaptive haptic rendering of a variable density 3D point cloud data 
at different levels of detail without pre-computing the mesh structure. We also incorporate features like rotation, translation 
and friction to provide a better realistic experience to the user. A proxy based rendering technique is used to avoid the pop-through 
problem while rendering thin parts of the object. Instead of a point proxy, a spherical proxy of variable radius is used which avoids 
the sinking of proxy during the haptic interaction of sparse data. The radius of the proxy is adaptively varied depending upon the local density of the point data 
using kernel bandwidth estimation. During the interaction, the proxy moves in small steps tangentially over the point 
cloud such that the new position always minimizes the distance between the proxy and the haptic interaction point (HIP). The 
raw point cloud data re-sampled in a regular 3D lattice of voxels are loaded to the haptic space after proper smoothing to 
avoid aliasing effects. The rendering technique is experimented with several subjects and it is observed that this functionality 
supplements the user's experience by allowing the user to interact with an object at multiple resolutions.
} 

\CCScatlist{\CCScat{Point cloud rendering, levels of detail, bounding surface, adaptive rendering, kernel bandwidth}}

\begin{document}


\firstsection{Introduction}

\maketitle

Haptic and visual rendering require the object geometry to be well defined. Generally, objects are 
represented by polygonal or triangular meshes as it makes further processing of rendering very simple. 
But these meshes have to be constructed and are not directly available. 
When an object with very fine details are represented by a triangular mesh, 
the number of small facets required increases drastically. The same object when displayed 
graphically as a whole, some facets may occupy less than one pixel, which results in a wastage of 
memory as well as the computation time. Hence when a highly detailed object is rendered haptically, 
the computational time can 
be reduced by reducing the number of small facets without affecting the haptic experience.
The scaling feature allows us to display the object using a fewer number of facets when 
it is less significant, \emph{i.e.} the object needs to be experienced at a gross
level and by more number of facets when it is to be experienced at 
finer details. With a mesh based representation, creating different levels of detail requires the mesh 
to be re-sampled and re-meshed. Re-sampling and re-meshing during the interaction 
is not possible for a dense data as it may not meet the required proxy update frequency for rendering haptically. 
For graphic rendering the ideal frame rate should be 30 frames per second, while to perceive the 
object haptically, proxy update frequency should be at least 1 KHz. Many of such issues 
can be resolved, if the object is represented as 
a point cloud. In the proposed method the average proxy update 
time is 0.03 ms which is 30 times faster than the required proxy update frequency.
Also there are devices like laser scanners, RGB-D cameras (eg. Microsoft Kinect) etc. 
that can directly create the point cloud data from a real object. Therefore the ease of 
data availability also makes rendering with point cloud preferable. A point cloud rendering 
algorithm uses the data directly and avoids the intermediate step of mesh construction. 
However the key issue is how to handle the proxy sinking problem in case of variable density 
point cloud. With a fixed radius, the proxy may fall through the point cloud of highly sparse 
region. If we use a proxy of sufficiently large radius, sinking problem can be solved but then 
the finer details cannot be experienced. In order to address this problem, the radius of a 
spherical proxy is adaptively changed by statistical bandwidth estimation of a Gaussian kernel.

In order to feel the 3D dense model at a closer level at run time we need to create hierarchies of 
levels of detail. With the Monge surface representation, such an hierarchy of levels of detail 
need to be pre-computed as given in \cite{cultural}. Authors have used a point proxy which is moved 
iteratively over the locally computed surface  to minimize the distance to the HIP during collision. Using 
this algorithm an user cannot rotate and translate the 3D object to render the rear part of object 
as depth data is generated here by using OpenGL depth buffer. This algorithm also does not incorporate 
the surface property like friction, which deters the realistic experience of user. In the proposed 
method a full 3D object is used instead of a Monge surface, where $z=f(x,y)$ and hierarchies of 
levels of detail are computed during haptic interaction time itself. There are very few haptic 
rendering algorithms for point cloud data. However, currently available algorithms cannot deal with 
scalability without constructing a surface or mesh of the whole point cloud. The Proxy based rendering 
is difficult with a 3D point cloud data as the standard bounding surface is not defined. So we propose 
to use a sphere of sufficient radius as a proxy to avoid the sinking in the point cloud data. 
To the best of the authors' knowledge, there exists no method for rendering a variable density point cloud at run 
time when it is being scaled, rotated and translated. Our proposed work facilitates the user with all these features and also 
incorporates surface friction.

Depending upon the user's interest the point cloud can be transformed (scaled, rotated or translated) prior to sampling. 
In our method, we re-sample the point cloud data in a regular 3D lattice of voxels and 
load them into the haptic space. In order to avoid the aliasing we use a mean 
filtering before sampling the 3D lattice. Since we do not have lattice structure of 
data, we cannot use other filters like Gaussian filter for smoothing.
In order to scale, rotate and translate the 3D model, the data points are 
multiplied by a suitable transformation matrix and then mapping is done to the 3D lattice. 
The mapped 3D points decide the actual object geometry for haptic rendering. 

In real world every object gives the sensation of a resistive force while touching it.
Without incorporation of  the surface friction in the algorithm \cite{cultural}, rendering gives a very slippery experience. 
In the proposed work, we simulate the surface friction to provide a more realistic experience to the user. 
Generally friction is simulated based on the local texture of the surface.
Since we are rendering point cloud data, we do not have any texture information. Therefore 
to provide the feeling of friction, we simulate it by slowing down the movement of proxy. 

The paper is organized as follows. Section \ref{liter_scale} gives an overview of relevant literature. 
In section \ref{PROPOSED METHOD}, we describe the proposed method of rendering a variable density point cloud data. This section also 
includes the key contributions of the paper, which describes how to handle 
scale change for zooming in and out, rotation and translation. 
Section \ref{RENDERING}  explains the force  and friction computation during the haptic interaction.
Results are presented in section \ref{RESULTS}. We draw the conclusions in section \ref{CONCLUSIONS}.

\section{LITERATURE REVIEW}\label{liter_scale}
In haptic rendering literature there are very few approaches with point cloud 
data. The common practice is to first create the surface or mesh from point 
data, and then render the mesh haptically. A good description of the basic haptic rendering using
such a mesh is provided in~\cite{Haptic_Intro}. In order to render the polygon based model, 
mainly two operations are involved: haptic collision detection and haptic force computation. 
There are several algorithms to detect the collision with high density points \cite{Coll_Detec}, 
implicit functions \cite{Implicit}, and polygonal mesh models \cite{Polyhedron_coll}. The most commonly used 
method is based on the polygon meshes which requires the calculation of the penetration depth 
of haptic interaction point(HIP) into the surface. If $\mathbf{D}$ is the depth of penetration in the model, the reaction 
force can be computed as $\mathbf{f}_r=-K\mathbf{D}$, where $K$ is the stiffness constant of 
the surface. Nevertheless, haptic rendering with polygon based model suffers from the problems 
such as dependency on the primitive level inputs like vertices and surface normal. 
The second problem associated with this rendering technique is ambiguity in determining the 
appropriate direction of the force while rendering thin objects. These problems have been solved 
by Zilles and Salisbury \cite{God_Obj}, and Ruspini \emph{et al.} \cite{Ruspini} by introducing the 
concept of god-object and the proxy point, respectively. The problem with the mesh based 
representation is that when the object is not fully enclosed by the bounding planes, a small hole 
may remain  and therefore the proxy sinks during the rendering process. Mesh based rendering is 
not amenable to render the object at different levels of detail as with a large number of polygons 
it requires to pre-compute the mesh at each level.

Another alternative rendering technique is the voxel based haptic rendering \cite{Vox_Samp}, \cite{Vox_Pointshell}. In these methods, 
properties of the object like stiffness, color, density and distance field  are sampled 
and stored in the 3D grid. The major drawback of this rendering technique is that we need 
to have a pre-computed distance field and computing distance field of dense model is itself 
a very complex and time-consuming process. This technique also suffers from the thin object 
problem. As regards rendering of a  point cloud data, Leeper \emph{et al.} discuss a constraint based approach of rendering where 
the points are replaced by spheres or surface patches of approximate size \cite{point_cloud3}. 
Another algorithm is described in  paper \cite{point_cloud2}, where author uses axis aligned bounding cube 
to fill the voids between the points and then use god object rendering technique. 
Haptic rendering with point cloud data has also been proposed by Lee \emph{et al.} in \cite{point_cloud}, which 
creates the moving least square surface from the point cloud data and then minimizes the distance between proxy and HIP. 
As this algorithm does not keep track of HIP penetration, it suffers from the thin object problem as well as 
creating a moving least square surface increases the complexity of the algorithm. We try to overcome these limitations by introducing a novel approach as explained in section 3.

\section{PROPOSED METHOD}\label{PROPOSED METHOD}
For better understanding, let us first assume that the point density of the object is uniform and this assumption will be relaxed later.
In order to haptically render the point cloud data we need to find out the penetration 
depth of HIP into the surface and the surface normal to apply force feedback. By taking these requirements 
into consideration, our algorithm tries to move the proxy on the surface in small steps in the tangential 
direction during the interaction and the new position always minimizes 
the distance between HIP and proxy which is the measurement of the amount of force to be fed back. 
\begin{figure}
\centering
\includegraphics[width=0.4\textwidth]{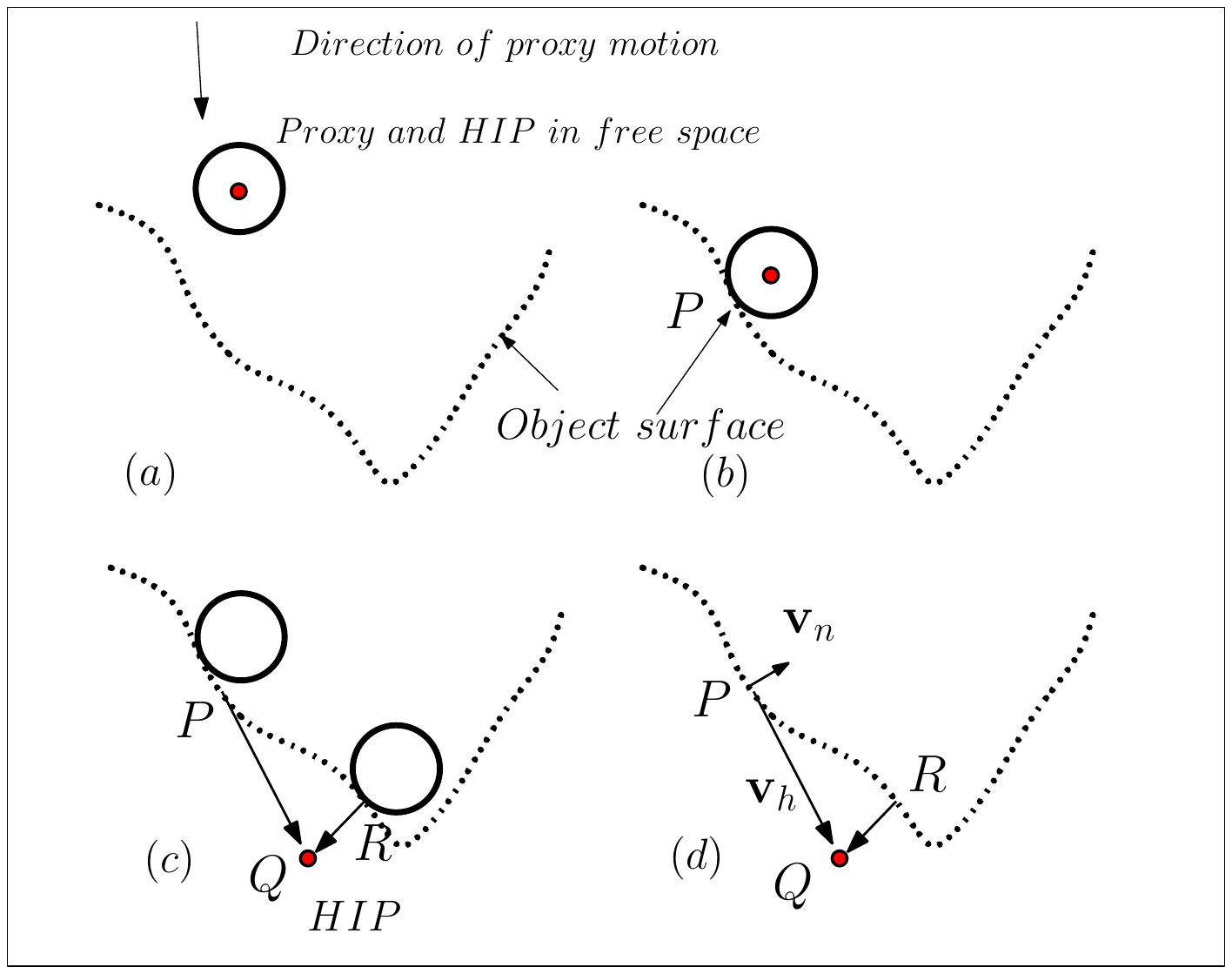}
\caption{Illustration of the proposed method to find the desired proxy position.}
\label{algorithm}
\end{figure}

The proposed rendering technique is in some sense similar to the technique given in \cite{symp}. 
For better understanding, let us consider a point data in 2D as shown in Fig. \ref{algorithm}. The small 
red  circle represents the HIP and the big black circle represents the proxy. In free space HIP and proxy move 
together and are  collocated as shown in Fig. \ref{algorithm}a. Let HIP and proxy 
touch the boundary at point $P$ as shown in Fig. \ref{algorithm}b. After this 
contact, HIP continues to move through the point data, and the proxy stays over the 
surface. In Fig. \ref{algorithm}c HIP has penetrated the surface and is inside the  
point cloud data and proxy remains over the surface. Once the HIP penetrates the 
points cloud data, reaction force need to be fed back to the user. In order to calculate the reaction force, 
penetration depth and direction of normal need to be computed. We calculate the  
penetration depth from the nearest point 
on the object boundary. We find the most appropriate position of proxy through 
our algorithm at $R$ as shown in Fig. \ref{algorithm}d. At point $R$ the vector 
joining the HIP and proxy is oriented opposite to the normal vector at $R$, 
hence the distance between HIP and subsequent proxy position is minimum at 
this point. The normal at the proxy location $P$ is shown with the vector $\mathbf{v}_n$ and the 
vector $\mathbf{v}_h$ is the vector to HIP from proxy. If $\mathbf{C}_p$
denotes the proxy center and $\mathbf{H}_p$ denotes the HIP point, the vector
$\mathbf{v}_h$ is given by
$\mathbf{v}_h$=$\mathbf{H}_p-\mathbf{C}_p$.
We move the proxy along the direction tangential to the surface and lying in the plane found 
by the vectors $\mathbf{v}_h$ and $\mathbf{v}_n$ to find the most appropriate 
location $R$ during the collision. The following equation is used to evaluate 
the tangent vector $\mathbf{v}_t$ at every step.  

\begin{equation} \label{tangeny_vector}
\mathbf{v}_t = \mathbf{v}_h-(\mathbf{v}_n.\mathbf{v}_h)\mathbf{n},
\end{equation} 
where $\mathbf{v}_n$ and $\mathbf{n}$ denote the surface normal  and unit surface normal at the point of contact of proxy with the surface.
With point cloud data we do not have other information about underlying objects 
such as surface normal or bounding surface. In our work we estimate the 
normal at the proxy location by using the points inside the spherical proxy as explained below.

In free space, there are no data points inside the spherical proxy as shown 
in Fig. \ref{algorithm}a. If points are present inside the proxy, we need to restrict the further 
movement of proxy into the object. Once the proxy touches the point cloud data, 
a function, $\mathbf{v}_n$ is computed dynamically which measures the sinking of proxy into the object \cite{symp}. 

\begin{equation}\label{normal_eq}
\mathbf{v}_n = \sum_{i=1}^N\mathbf{d}_i.
\end{equation}

The function $\mathbf{v}_n$ can be computed from the radial overshoot 
of each point as given in equation \ref{normal_eq}. For each point $\mathbf{p}_i$, overshoot $\mathbf{d}_i$ can 
be evaluated using equation \ref{overshoot}, where $r_p$ and $\mathbf{C}_p$ are 
the radius and the center of the proxy, 
respectively. 
\begin{equation}\label{overshoot}
\mathbf{d}_i=(r_p-|{\mathbf{C}_p-\mathbf{p}_i}|)\frac{(\mathbf{C}_p-\mathbf{p}_i)}{|\mathbf{C}_p-\mathbf{p}_i|}.
\end{equation}

The unit surface normal as used in equation \ref{tangeny_vector}, at the proxy point during the interaction is given by
\begin{equation}\label{normal}
\mathbf{n} = \frac{\mathbf{v}_n}{|\mathbf{v}_n|}.
\end{equation}

The vector $\mathbf{n}$ is estimated at each iteration and it gives the direction 
along which proxy should be moved to reduce sinking.

\subsection{Proxy Update Algorithm}
The proposed algorithm can be illustrated using the following steps.
\begin{enumerate}
\item[1] Let us consider the situation when HIP is outside the object.
The moment proxy starts sinking into the point cloud,   
the proxy may enclose some points and $\mathbf{v}_n$ 
becomes non-zero. To avoid the proxy from further sinking, we move the proxy 
by an amount of $k_n$ in the normal direction $\mathbf{v}_n$.
If $(|\mathbf{v}_n|>\zeta)$  where $\zeta$ is a small non-zero value then the proxy position is updated according to the 
following equation.
\begin{equation}\label{move_normal}
\quad \mathbf{C}_p\leftarrow \mathbf{C}_p+k_n\,\mathbf{v}_n. 
\end{equation}
In our work the value of $\zeta$ is set 0.05 times the proxy radius.
\item[2] Once the HIP penetrates the object,  the value of  $(\mathbf{v}_n.\mathbf{v}_h)$ becomes negative and the 
magnitude of $\mathbf{v}_t$ becomes non-zero from equation \ref{tangeny_vector}. Then the proxy is updated along the 
tangential direction according to the following equation.

\begin{equation}\label{move_tangential}
\quad \mathbf{C}_p\leftarrow \mathbf{C}_p+\delta\,\mathbf{v}_t+k_n\,\mathbf{v}_n.
\end{equation}
The value of $\delta$ decides the step size to move the proxy in the tangential 
direction. If the proxy is updated only along the tangential direction, there exists a possibility 
that the proxy may sink into the point cloud. To avoid this we incorporate the term $k_n\,\mathbf{v}_n$ in the proxy 
movement which moves the proxy along the normal direction by a small amount. The value of ${k_n}$ is very small and 
it is chosen such that proxy neither leaves the surface nor remains inside the surface.

\item[3] When HIP comes in free space, the proxy is expected to latch on to HIP as explained earlier. But sudden 
withdrawal in penetration of the HIP toward the free space may lead to sudden vibration.  
To ensure a smooth movement of the proxy during withdrawal process, the following proxy update equation is used. 
\begin{equation}\label{move_withHIP}
\quad \mathbf{C}_p\leftarrow \mathbf{C}_p+k_h\,\mathbf{v}_h,
\end{equation}

\end{enumerate}
To ensure all three conditions, we run a separate thread to check the magnitude of 
$\mathbf{v}_t$ and $\mathbf{v}_n$ continuously. 
The parameters $k_h$, $k_n$ and $\delta$ are chosen empirically 
and are less than 1. In this study we have used $k_h$=0.002, $k_n$=0.064 and $\delta$=0.000008. For a larger value of $\delta$, proxy converges faster but it creates 
some undesirable vibration in the haptic device. On the other hand for a
smaller value of $\delta$, it does not cause vibration but convergence 
is a bit slower. It may be noted that the normal can also be computed locally from point cloud data 
by fitting the surface points. But it cannot  measure the sinking of proxy,  
which is needed to keep the proxy on the point cloud data and hence such a technique cannot 
be used to update the proxy.

\subsection{Kernel Density Estimation}
The above discussed algorithm assumes that the point cloud density is uniform over the data.
In fact the density of points and the details of the object are closely related. For example 
an object with a flat surface requires less number of points whereas an object with very fine
details requires a large number of points. Because of these reason the point cloud density 
over the surface of an arbitrary object cannot be considered uniform. 
 
In our work we use  a variable proxy radius depending on the local point cloud density
using kernel bandwidth estimation. The kernel density estimator from an unknown distribution of the samples $(x_1,x_2,..x_n)$
is given by  
\begin{equation}\label{pdf_eq}
\hat{f_b}(x)=\frac{1}{nb}\sum_{i=1}^{n}K(\frac{x-x_i}{b}),
\end{equation}
where the parameter $b$ is called the bandwidth and $K(.)$ is the chosen 
kernel. The function $\hat{f_b}$ estimate the $pdf$ of the distribution of the sample with a given 
bandwidth $b$ as explained in ~\cite{Rayk06}. Proxy radius is varied 
according to the relation $r_p=\beta{b}$ where $\beta$  is an appropriate scale factor. The proxy variation is limited by 
two limiting radii $r_{1}$ and $r_{2}$ where $r_{1}$ denote the lowest possible proxy radius and $r_{2}$ denote the highest 
possible proxy radius. Assuming a Gaussian kernel, the optimum bandwidth $b$ for a univariate data is given by equation \ref{bw_eq}.
\begin{equation}\label{bw_eq}
b=1.06\hat\sigma^{-1/5},
\end{equation}
where $\hat\sigma$ denotes the standard deviation (SD) of the sample points. 
The proxy radius is varied according to the relation given in equation \ref{bw_eq} between the two limiting radii $r_1$
and $r_2$. We estimate $\hat\sigma$ locally over a neighborhood from the given point cloud.

\subsection{Scaling, Rotation and Translation of the Object}
In order to find out the normal to the surface as explained in the previous 
section, we need to find out the points inside the proxy. To reduce the 
search space for finding neighborhood point of the proxy, we partition 
the haptic space using a regular 3D lattice of voxels and map the point cloud data to 
this 3D lattice. 
 
\begin{figure}
\centering
\includegraphics[width=0.2\textwidth]{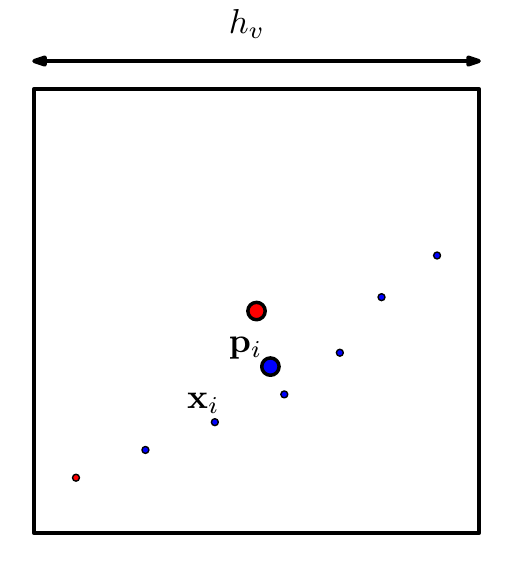}
\caption{Illustration of data smoothing for re-sampling on a discrete grid. The small blue points represent the point 
cloud data and the big blue circle $\mathbf{p}_i$ represents the mean of all the points inside that grid. The 
red circle denotes the center of grid.}
\label{mean_fig}
\end{figure}

We first filter the point cloud data with a mean filter before re-sampling them 
to the 3D lattice, to avoid the bump during rendering.
Let us consider a single voxel as shown in Fig. \ref{mean_fig}. 
Let $\{\mathbf{x}_1, \mathbf{x}_2, ..,\mathbf{x}_m\}$ be the unordered data points inside the lattice, 
and the mean of $\mathbf{x}_i's$ is given by
\begin{equation}\label{mean}
\mathbf{p}_i=\frac{1}{m}\sum_{i=1}^{m}\mathbf{x}_i.  
\end{equation} 
The mean is calculated dynamically at run time. This prevents unnecessary memory requirements and 
enhance the speed of execution.  
We represent the object only by the mean of points inside each lattice 
instead of using all the points inside the lattice. The lattice which has data 
points and subsequently the mean of points is called an active lattice. 
We observe the proposed technique  helps reducing 
the run time with a good  rendering experience. 

One of the key contributions of this work is to provide the feature like scalability 
which enables the user to change the resolution of object graphically as well 
as haptically. To get the real sense of overall structure one needs to scale up 
the object to feel the fine details. If a highly dense model is displayed as whole, 
we cannot feel the minor variation of the surface. 
In practice, users need to experience objects of all sizes at different resolutions to get
a feel of overall structure to a finer details from the same data set.
For instance, if a user wants to feel the 3D point cloud model of an entire monument, 
one needs to scale down the model to fit it into the 3D lattice. However, if the user 
wishes to feel a smaller part of the monument like a pillar at a closer level, that part of the model has to be scaled 
up appropriately. To incorporate all three features scaling, rotation and translation, we use an affine transform 
given by equation \ref{transformation_eq}, where matrix $T$ can be a scaling or a rotation matrix and  $T_t$ is a 
translation vector. 
\begin{equation}\label{transformation_eq}
\mathbf{x}_{Ti}={[T]}{[\mathbf{x}_i]}+{[T_t]}.  
\end{equation}

To modify the resolution of the object, we transform the data points by using a scaling matrix $T$ which 
can be any diagonal matrix. After scaling up, the transformed points $\mathbf{x}_{Ti}s$  become dispersed 
and may require more number of lattices to inhabit. In this way, we represent the scaled object with more data points 
which increases the graphic and haptic resolution. Similarly while scaling down, number of active grids and hence the 
resolution decreases. As explained earlier, data smoothing is performed 
on the transformed points after which they are mapped back into the 3D lattice.
The sampled set of points, $s(\mathbf{p})$ are finally rendered haptically. 

Rotation allows the user to feel even the rear side of 
the object. We observed that, rotation feature  gives the user a better perception about the overall structure. In 
order to rotate the model,  we transform the point cloud data by using an appropriate orthonormal rotation matrix.  
The rotated points are then mapped back  into the lattice after proper smoothing. 
Since our haptic workspace is limited, we cannot render a scaled up version of whole object at a time. 
In such cases, we use the translation vector $T_t$  to bring the desired part of object to the 
haptic workspace. In this way an user can feel the complete 
object more precisely at a closer level when needed and zoom out when context is desired.

\section{RENDERING}\label{RENDERING}
To render the reaction force we need the 
penetration depth of HIP into the surface. 
If $(\mathbf{v}_n.\mathbf{v}_h)<0$ then proxy has touched the object and a force 
needs to be fed back by the haptic device. Subsequently, the reaction force is 
evaluated as $\mathbf{f}_r=-K\mathbf{D}$, where $K$ is the Hooke`s constant and 
$\mathbf{D}$ is the penetration depth of HIP into the object. For finding the 
penetration depth $\mathbf{D}$ we assume the sinking of the proxy into the point 
cloud negligible. Hence we approximate $\mathbf{D}$ as
\begin{eqnarray}\label{penetration_depth}
\mathbf{D} &=& (|\mathbf{v}_h|-r_p)\frac{\mathbf{v_h}}{|\mathbf{v_h}|}
\hspace{0.4cm}  for \hspace{0.4cm} |\mathbf{v_h}|\ge{r_p} \nonumber\\
     &=& \mathbf{0} \hspace{0.4cm}  otherwise ,
\end{eqnarray}
where $r_p$ is the radius of the proxy which is varied according to the equation \ref{bw_eq}.

\subsection{Surface Friction}
\begin{figure}
\centering
\includegraphics[width=0.28\textwidth]{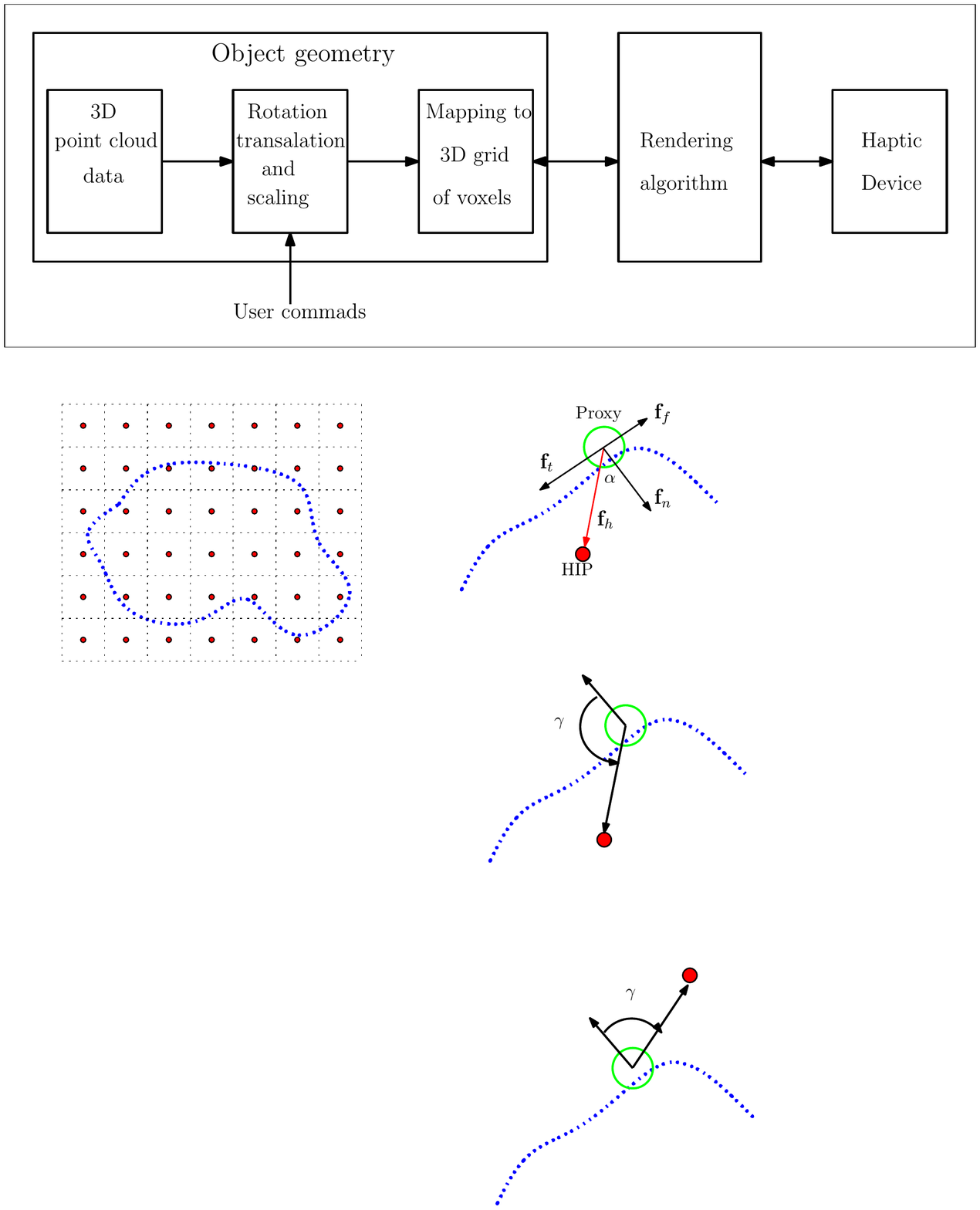}
\caption{Illustration of calculating resultant force along the tangential direction 
for the proxy movement after incorporating surface friction.}
\label{surface_friction}
\end{figure} 
As mentioned earlier, we simulate friction in our method to provide a more realistic feeling of the 
surface. We try to incorporate the friction by slowing down the proxy movement. As shown in 
Fig. \ref{surface_friction}, $\mathbf{f}_h$ is the force in the direction of $\mathbf{v}_h$. 
The normal component of the force is $\mathbf{f}_n$ which is equal to $|\mathbf{f}_h|\cos\alpha$. 
Similarly the tangential component of the force is $|\mathbf{f}_h|\sin\alpha$. 
To move the proxy over the surface, a static frictional force must be overcome by the 
applied force. When enough force is applied, proxy starts the motion and 
a transition occurs from static to dynamic friction state. The magnitude of 
static friction is proportional to $\mathbf{f}_n$.
\begin{equation}\label{static_friction}
\mathbf{f}_s=\mu_s |\mathbf{f}_n|.  
\end{equation}
Here $\mu_s$ denotes the static friction coefficient. Once the applied 
force exceeds the static friction force, dynamic friction force occurs against 
the motion of proxy. The magnitude of dynamic friction force is $\mu_d|\mathbf{f}_n|$, 
where $\mu_d$ is coefficient of dynamic friction force and is usually less than $\mu_s$. Let $\mathbf{f}_t$ be  
the force  along the tangential direction. Then the resultant force $\mathbf{f}_r$ 
on the proxy in the tangential direction is given by $|\mathbf{f}_t|-\mu_d|\mathbf{f}_n|$.\\
\begin{eqnarray}\label{resultant_force}
\mathbf{f}_r &=&\mathbf{f}_t (1 -\mu_d\cot\alpha )\hspace{0.4cm} if \hspace{0.4cm} |\mathbf{f}_t|\ge\mu_s|\mathbf{f}_n|. \\ \nonumber          
\end{eqnarray}
For  $|\mathbf{f}_t|$ less than $\mu_s |\mathbf{f}_n|$, there is no proxy movement 
as expected. Equation \ref{resultant_force} provides a direct description of friction 
force. However, for haptic rendering, such an equation has to be absorbed in defining 
the proxy movement by slowing  it down. In equation \ref{move_tangential}, $\delta$ 
is the factor which decides the speed of movement of the proxy along the tangential 
direction. Comparing equation \ref{move_tangential} and \ref{resultant_force}, we 
observe that $\mathbf{f}_r$ is the net force responsible for the movement of proxy 
in the tangential direction and is proportional to $\mathbf{v}_t$. Hence the 
parameter $\delta$ with friction can be given by $\delta_f=\delta(1-\mu_d\cot\alpha)$. 
We modify equation \ref{move_tangential} as 

\begin{equation}
\mathbf{C}_p^{(k+1)}=\mathbf{C}_p^{(k)}+\delta(1-\mu_d\cot\alpha^{(k)})\mathbf{v}_t^{(k)}.
\end{equation}
In this way, we reduce the step size of the proxy movement to provide the feeling of friction. 

\begin{figure*}
\centering
\subfloat[]{\label{result_scalinga}
\includegraphics[width=0.22\textwidth]{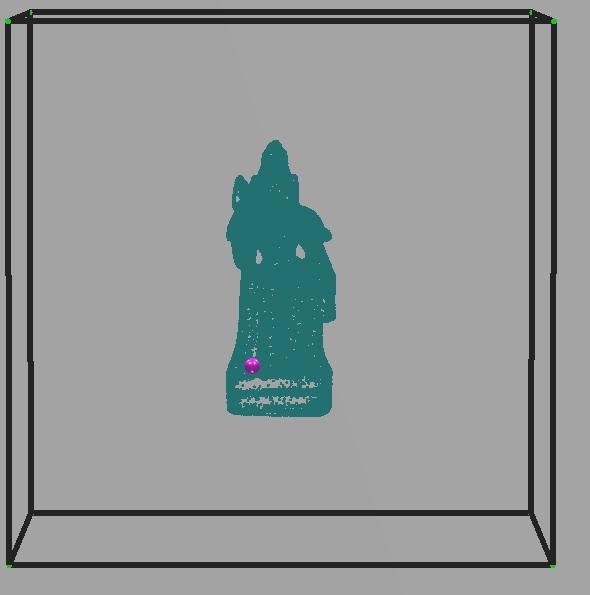}}
\subfloat[]{\label{result_scalingb}
\includegraphics[width=0.22\textwidth]{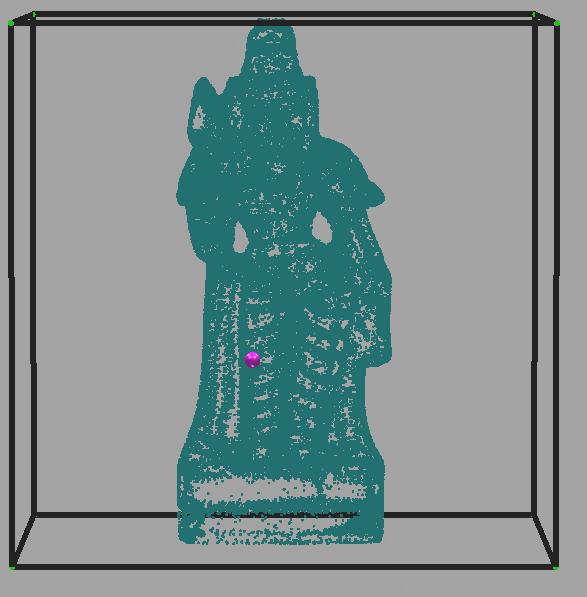}}
\subfloat[]{\label{result_scalingc}
\includegraphics[width=0.22\textwidth]{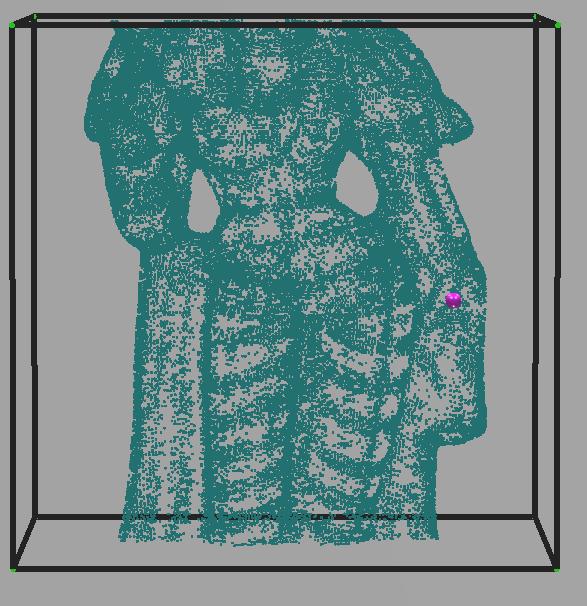}}
\caption{A point cloud data of an Indian idol rendered at 3 different resolutions. (Data Courtesy: \emph{www.archibaseplanet.com})}
\label{result_scaling}
\end{figure*} 

\begin{figure}
\centering
\subfloat[]{\label{result_rota}
\includegraphics[width=0.15\textwidth]{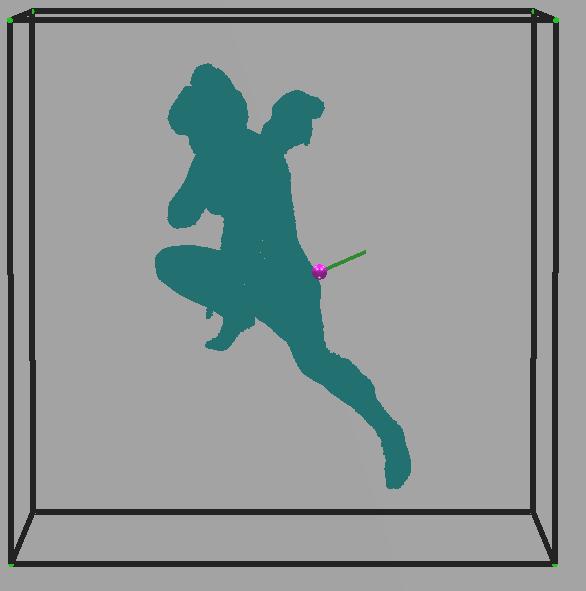}}
\subfloat[]{\label{result_rotb}
\includegraphics[width=0.15\textwidth]{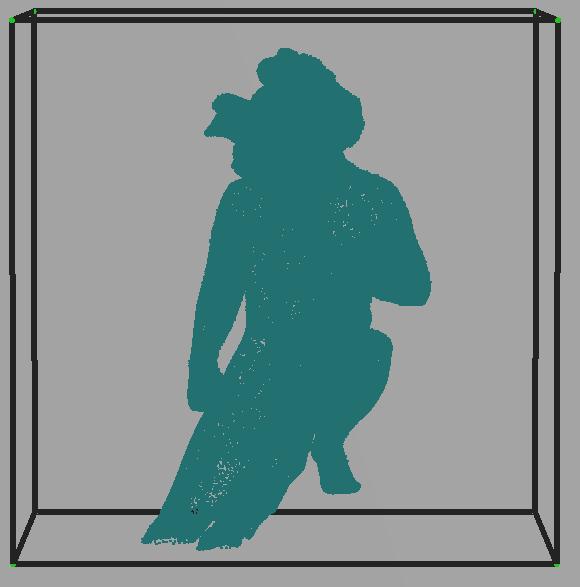}}
\subfloat[]{\label{result_rotc}
\includegraphics[width=0.15\textwidth]{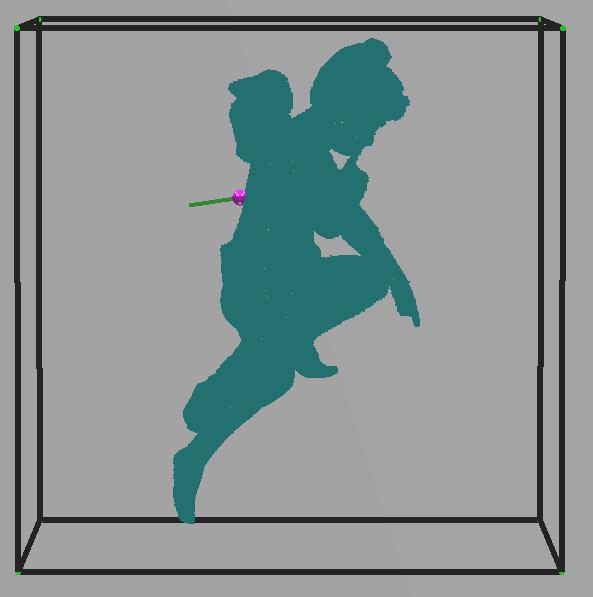}}
\caption{Point cloud corresponding to an angel rendered at 3 different resolutions. a) At lowest level of detail b) Model rotated by 90 degree and at higher 
level of detail c) Model rotated by 180 degree along with scaling. (Data Courtesy: \emph{www.archibaseplanet.com})}
\label{result_rot}
\end{figure} 

\begin{figure}
\centering
\subfloat[]{\label{result_rotationa}
\includegraphics[width=0.15\textwidth]{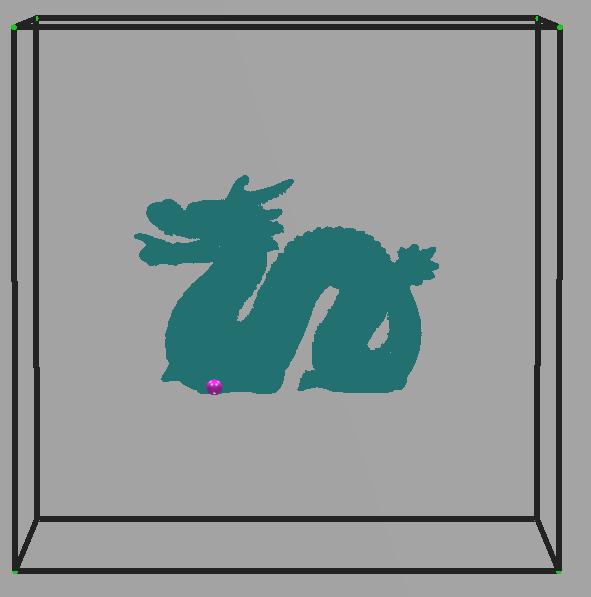}}
\subfloat[]{\label{result_rotationb}
\includegraphics[width=0.15\textwidth]{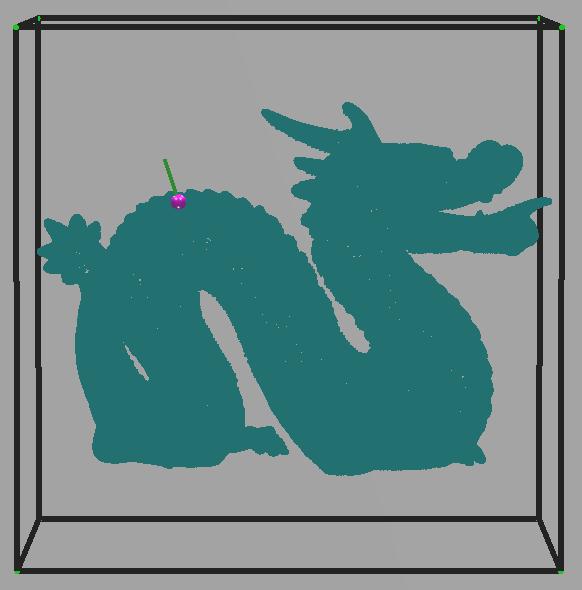}}
\subfloat[]{\label{result_rotationc}
\includegraphics[width=0.15\textwidth]{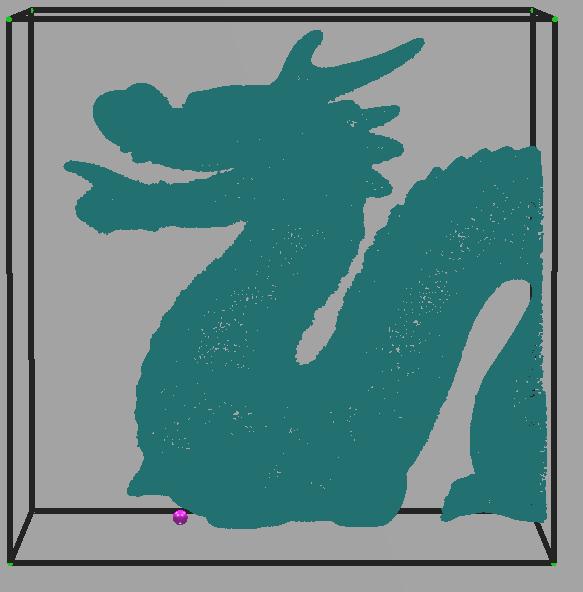}}
\caption{a) Point cloud data correponding to a Dragon at the lowest resolution b) Model of the dragon scaled and rotated by 180 degree and 
c) Model of the dragon scaled up and translated to bring the front part of it into the active cube. 
(Data Courtesy: \emph{www.archibaseplanet.com})}
\label{result_rotation}
\end{figure} 

\begin{figure} 
\centering
\includegraphics[width=0.4\textwidth]{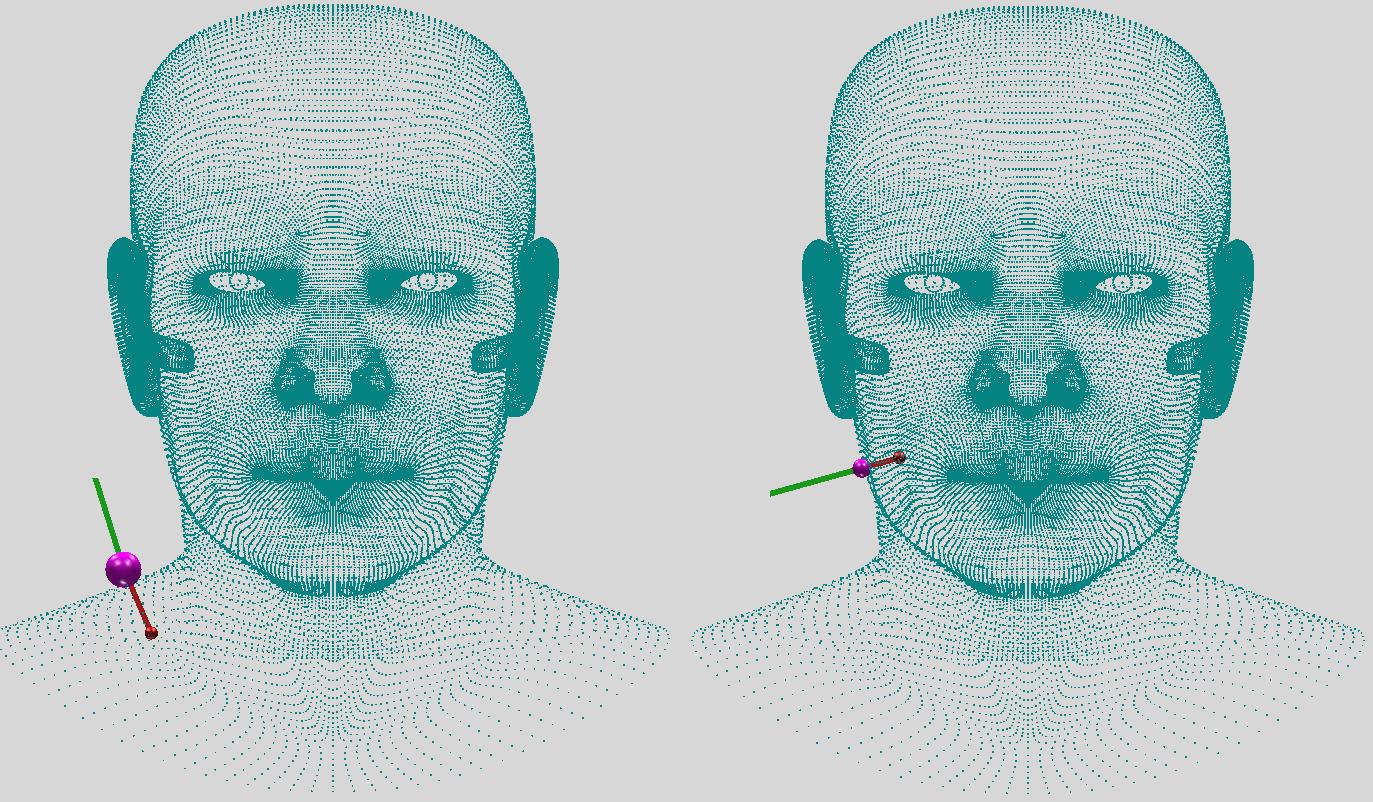}     
\caption{Illustration of estimated varying radius of proxy at different density locations. (data courtesy:- www.turbosquid.com)}
\label{bonz_fig} 
\end{figure}

\section{RESULTS}\label{RESULTS}
The proposed method was implemented using visual C++ in a Windows XP platform with 
CORE 2QUAD CPU @2.66 GHz with 2 GB RAM. A Falcon haptic device from NOVINT with 10 $cm^3$ 
cube of interaction space is used to feel the reaction force. 
The 3D lattice of size  $300\times300\times300$ is used to map the point cloud 
data  with a lattice spacing of 0.04 m between the adjacent lattice nodes. 
We can increase the resolution by decreasing the space between two adjacent 
lattice. We use HAPI library for haptic rendering. Fig. \ref{result_scaling} shows the 
point cloud corresponding to an Indian idol with 47516 points. The proxy is shown in pink color
in the figure. Fig. \ref{result_scalinga} corresponds to the lowest  
level of surface details. We successively zoom the object and the corresponding scaled objects 
are shown in Fig. \ref{result_scalingb} and Fig. \ref{result_scalingc}. It is quite clear from the 
figure that later figures have more details and a visual perception of 
closeness. In the above and subsequent two figures, haptic interaction space is shown by a black wire mesh. The green line attached to 
the proxy shows the normal direction computed at the shown proxy location. 
Fig. \ref{result_rot} demonstrates the rotation feature. Fig. \ref{result_rota} is shown at the 
lowest resolution and it allows rendering at a gross level, while in Fig. \ref{result_rotb} and Fig. \ref{result_rotc}, it is  
rotated by 90 and 180 degree, respectively. This feature becomes desirable when the  user 
is interested to feel the rear part of the object. Fig. \ref{result_rotationa} and \ref{result_rotationb} are other examples 
of scaling and rotation. The scaled up version of Fig. \ref{result_rotationa} is shown in Fig. \ref{result_rotationc}, in which an user  
can feel even minute variations of the model. Hence they can have a more realistic experience. 
Whenever the scaled object goes out of the haptic space, it gets clipped off. Fig. \ref{result_rotationc} 
demonstrates the translation feature in which the object is translated to bring the front part of it into an active cube and 
feel it more precisely. Fig. \ref{bonz_fig} demonstrates two different regions of point cloud model having varying density and 
the computed surface normals at corresponding locations. The left side of the Fig. \ref{bonz_fig} shows the low density region 
where the radius of proxy is large and the right side of the Fig. \ref{bonz_fig} shows the higher density region and here the radius of proxy is smaller. 
The average time required to compute the normal vector 
is 0.0328 ms, which means we could update the proxy nearly 30 times faster 
than the required haptic update frequency of 1 KHz. The average time required 
for data smoothing and loading it into haptic cube depends on the density of 
input point cloud data and it was observed to be  around 1.109 s for a point cloud data 
with 543652 points.

The fact that the natures of the two plots
are quite different during the interaction demonstrate that
the reaction force does not get affected due to the change in
proxy radius.

\begin{figure}
\centering
\includegraphics[width=0.5\textwidth]{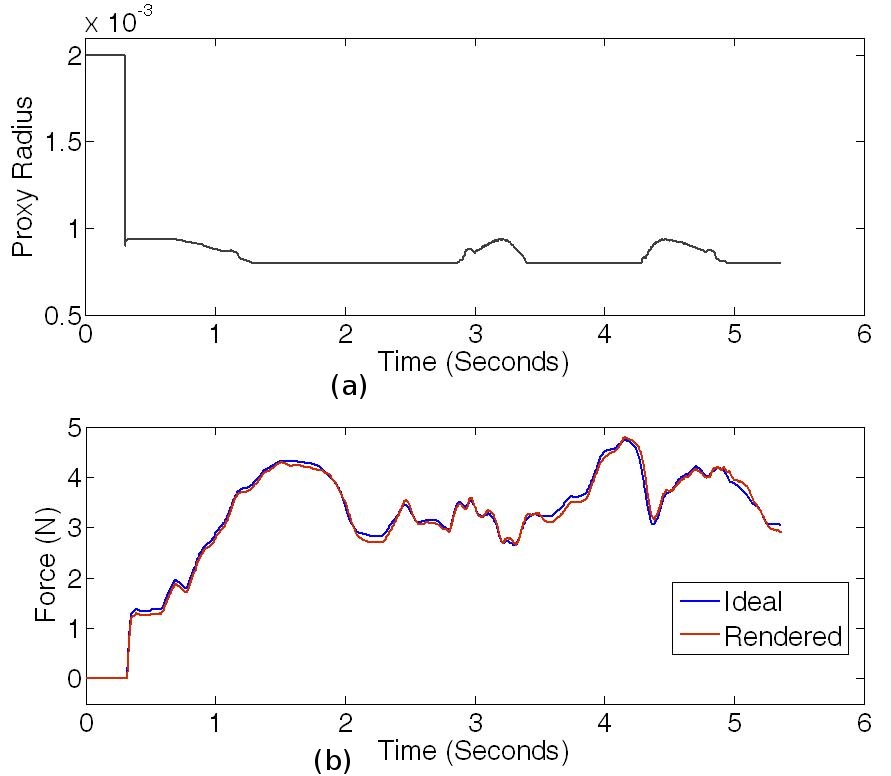}
\caption{(a) Variation of proxy radius in meter during haptic interaction 
with a spherical object of radius 0.025 m. (b) comparison of rendered force with actual one during the interaction.}
\label{validation1}
\end{figure}

\begin{figure}
\centering
\includegraphics[width=0.5\textwidth]{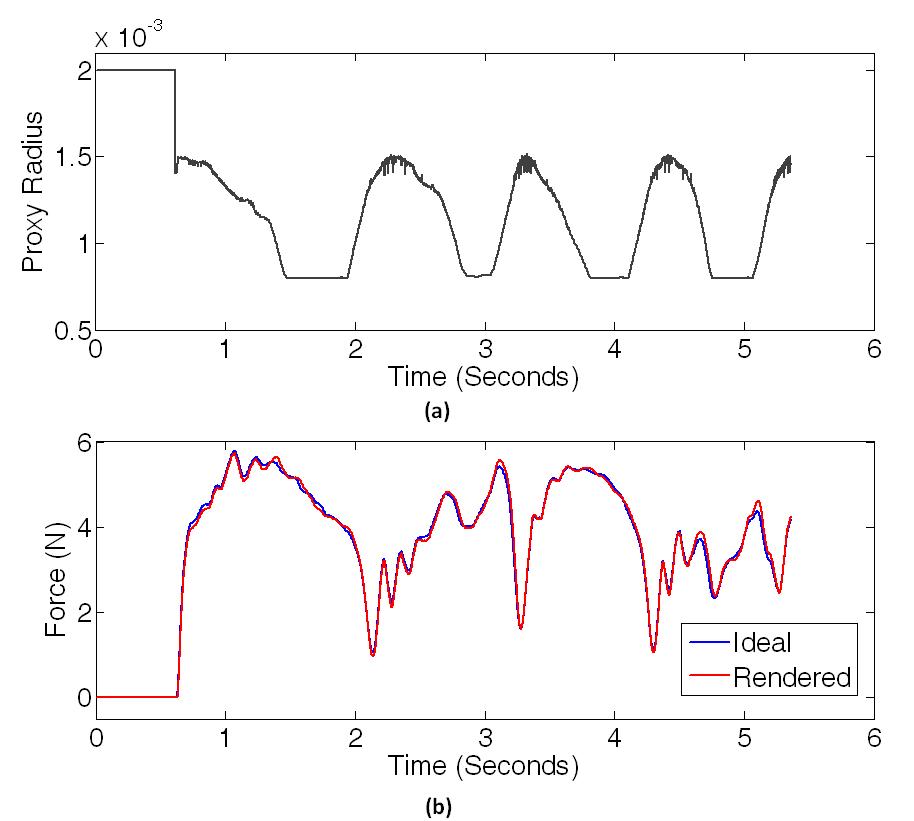}
\caption{(a) Variation of proxy radius in meter during haptic interaction with a scaled up sphere of radius 0.04 m.
(b) Comparison of rendered force with the actual one.}
\label{validation2}
\end{figure}

In haptic rendering there is no standard way to validate the results.   
We demonstrate the validation of our results by using a point cloud model of sphere having a radius of 0.025 m. 
The illustration of variation of proxy radius during haptic interaction with the object is shown by Fig. \ref{validation1}(a). 
In free space proxy radius is 0.025 m. As the proxy touches the point cloud model of object, its radius changes according to the local 
density of points. We also computed the ideal reaction force by using the known implicit equation of 
the sphere. Force computed by using the implicit surface equation is compared 
with the reaction force computed by the proposed rendering technique.
From Fig. \ref{validation1}(b) we can see that the ideal reaction force is quite close 
to the rendered force. In order to validate the scaling technique,  
we scaled up the model of sphere and recorded the variation of proxy radius and rendered force. Fig. \ref{validation2}(a) and 
Fig. \ref{validation2}(b) illustrate the variation of proxy radius and the comparison between the rendered and the ideal force 
at a scaled up level respectively. We noticed that there is no dependency between the reaction force and proxy radius. 
For experiments related to Fig. \ref{validation1} and  \ref{validation2}, $\mu_d$ is set to zero. For a non-zero value of $\mu_d$, 
the computed force is found to lag the actual force, as is expected due to presence of friction. 
We observed that, even after scaling the sphere, 
error was within the limit of 0.4$\%$ of the rendered force.  Further, we asked 
different subjects to evaluate the performance of the proposed technique on various 3D point cloud data. 
The subjects reported a very good experience and did not complain about any unusual slipperiness or vibration.

\section{CONCLUSIONS}\label{CONCLUSIONS}
In this work, we proposed an algorithm to render a variable density point cloud data without 
constructing the surface or polygonal mesh out of them. Our primary goal 
was to efficiently render the 3D point cloud data with attributes like scaling, 
rotation, translation, and friction to augment the users' experience. The distance between the proxy and HIP was minimized iteratively 
during haptic interaction. The average proxy update time was 0.0328 ms even 
for a highly dense point data which is far less than the required haptic update 
time of 1 ms. We also proposed a way to get rid of unnecessary vibration from 
haptic device due to mapping a large number of points into a small haptic active 
space. The method enhances the efficiency of searching neighborhood points 
around the proxy which subsequently makes the normal computation faster. 
In order to handle the variation in point density, we have proposed a run time kernel bandwidth estimation technique 
that allows the size of the proxy radius to change adaptively. 
We validated the proposed algorithm by using a point cloud model of a sphere and found 
the force obtained by the proposed rendering technique to be almost same as the ideal one. 
We are currently exploring the possibility of using an adaptive octree for partitioning of the haptic workspace.


\bibliographystyle{abbrv}
\bibliography{template}
\end{document}